\documentclass[twocolumn,floatfix,aps,superscriptaddress,showpacs]{revtex4}
\usepackage{amsmath}
\usepackage{graphicx}
\begin{document}
\title{Andreev reflection in ferromagnet--superconductor junctions}

\author{M. J. M. de Jong}
\affiliation{Philips Research Laboratories, 5656 AA  Eindhoven, The Netherlands}
\affiliation{Instituut-Lorentz, University of Leiden, 2300 RA  Leiden, The Netherlands}
\author{C. W. J. Beenakker}
\affiliation{Instituut-Lorentz, University of Leiden, 2300 RA  Leiden, The Netherlands}
\date{September 1994}
\begin{abstract}
The transport properties of a ferromagnet--superconductor (FS) junction
are studied in a scattering formulation.
Andreev reflection at the FS
interface is strongly affected by the exchange interaction
in the ferromagnet.
The conductance $G_{\text{FS}}$ of a ballistic point contact between F and S
can be both larger or smaller
than the value $G_{\text{FN}}$ with the superconductor in the normal state,
depending on the ratio of the exchange and Fermi energies.
If the ferromagnet contains
a tunnel barrier (I), the conductance $G_{\text{FIFS}}$
exhibits resonances which do not
vanish in linear response --- in contrast to the
Tomasch oscillations for non-ferromagnetic materials.
\end{abstract}
\maketitle

Electrons in a metal can not penetrate into a superconductor if their
excitation energy with respect to the Fermi level
is below the superconducting gap $\Delta$.
Still, a current may flow through a normal-metal--superconductor (NS)
junction in response to a small applied voltage $V < \Delta /e$,
by means of a scattering process known as
Andreev reflection \cite{and64}: An electron in the normal metal
is retroreflected at
the NS interface as a hole and a Cooper pair is carried
away in the superconductor.
Andreev reflection near the Fermi level conserves energy and momentum
but does not conserve
spin --- in the sense that the incoming electron and
the Andreev reflected hole occupy opposite
spin bands. This is irrelevant for materials with spin-rotation
symmetry, as is the case for normal metals.
However, the change in spin band associated with Andreev reflection may
cause an anomaly in
the conductance of (metallic) ferromagnet--superconductor (FS) junctions,
because the spin-up and the spin-down band in the ferromagnet are
different.
This paper contains a theoretical study of Andreev reflection in
FS junctions.
We use a scattering approach based on the Bogoliubov-de Gennes equation
to study the transport properties for zero temperature and small
$V$ ($e V \ll \Delta$).
We will concentrate on two distinct effects, which we think
are experimentally observable.
First, due to the change in spin band there is no
complete Andreev reflection at the FS
interface. This has a clear influence on the conductance and the
shot-noise power of clean FS point contacts. Second, the different
spin-up and spin-down wavevector at the Fermi level may lead to
quantum-interference effects. This shows up in the linear-response
conductance of FIFS junctions, where the ferromagnet contains an
insulating tunnel
barrier (I).

In the past, FS junctions
with an insulating layer between the ferromagnet and the
superconductor have been used
in spin-dependent tunneling experiments \cite{mes94}.
There the emphasis was on the voltage scale $e V \gtrsim \Delta$
and Andreev reflection did not play a role.
Tunneling through  S--Fi--S
junctions, where Fi is a magnetic insulator,
has been studied both experimentally
\cite{sta85} and theoretically \cite{dew85,kup91}.
In addition, there has been theoretical work on
the Josephson effect in SFS junctions \cite{bul82,kup86}.
An experimental investigation of the boundary resistance
of sputtered SFS sandwiches has also been reported \cite{fie90}.
The importance of phase coherence was
demonstrated in a recent experiment
\cite{pet94}, in which the effect of a remote
superconducting island on the conductance of a ferromagnet was observed.
We do not know of any previous theoretical work on
the influence of
Andreev reflection on the sub-gap conductance of an FS junction.

In order to clarify the effects we are aiming at, let us first give an
intuitive and simple description of the conductance through a ballistic
FS point contact.
A ferromagnet is contacted through a small area with a superconductor.
The transverse dimensions of the contact area are much smaller than the
mean free path and the interface is clean, so that
the conductance is completely determined by the scattering
processes that are intrinsic to the FS interface.
In a semiclassical approximation all scattering
channels (transverse modes in the point contact at the Fermi level)
are fully transmitted,
when the superconductor is in the normal state.
Let $N_\uparrow (N_\downarrow)$ be the number
of up(down)-spin channels, so that
$N_\uparrow \geq N_\downarrow$.
At zero temperature, the spin channels do not mix
and the conductance is given by the Landauer formula
\begin{equation}
G_{\text{FN}} = \frac{e^2}{h} \, ( N_\downarrow + N_\uparrow) \: .
\label{e1}
\end{equation}
In the superconducting state, the spin-down electrons of
all the $N_\downarrow$ channels are Andreev reflected into
spin-up holes.
They give a double contribution to the conductance since
$2e$ is transferred at each Andreev reflection.
However, only a fraction $N_\downarrow/N_\uparrow$ of the
$N_\uparrow$ channels can be Andreev reflected,
because the density of states in the spin-down band is smaller than
in the spin-up band.
Therefore, the resulting conductance is
\begin{equation}
G_{\text{FS}} = \frac{e^2}{h} ( 2 N_\downarrow +
2 \frac{N_\downarrow}{N_\uparrow} N_\uparrow ) =
4 \, \frac{e^2}{h} \, N_\downarrow  \: .
\label{e2}
\end{equation}
Comparison of Eqs.\ (\ref{e1}) and (\ref{e2}) shows that $G_{\text{FS}}$
may be either larger or smaller than $G_{\text{FN}}$ depending on the
ratio $N_\uparrow / N_\downarrow$.
If $ N_\downarrow / N_\uparrow < 1/3$ then $G_{\text{FS}} < G_{\text{FN}}$,
and vice versa. This qualitative argument can be substantiated by
an explicit calculation, as we now show.

For the conduction electrons inside the ferromagnet we apply
the Stoner model, using
an effective one-electron Hamiltonian with an exchange interaction.
The effect of the ferromagnet on the
superconductor is twofold. First, there is the influence of the exchange
interaction on states near the interface. This will be fully taken into
account. Second, there is the effect of the
magnetic field due to the magnetization of the ferromagnet.
Since this field --- which is typically a factor thousand smaller than
the exchange field ---
does not break spin-rotation symmetry it will
be neglected for simplicity.
Note that in typical layered structures
the magnetization is parallel to the FS interface, so that
it has no influence on the superconductor at all.

Transport through NS junctions has successfully been
investigated through the Bogoliubov-de Gennes equation
\cite{btk82,lam91,tak92,bee92}.
Here, we adopt this approach for an FS
junction.
In the absence of spin-flip scattering in the ferromagnet,
the Bogoliubov-de Gennes equation
breaks up into two independent matrix
equations, one for the up-electron, down-hole
quasiparticle wavefunction
$(u_\uparrow, v_\downarrow)$ and another one for
$(u_\downarrow, v_\uparrow)$. Each matrix equation has the form \cite{SM&A}
\begin{equation}
\left( \begin{array}{cc}
{\cal H}_0 - h
& \Delta \\
\Delta^* &
- ( {\cal H}_0 + h )
\end{array} \right)
\left( \begin{array}{c}
u_\uparrow \\ v_\downarrow
\end{array} \right)
=   \varepsilon
\left( \begin{array}{c}
u_\uparrow \\ v_\downarrow
\end{array} \right)
\: .
\label{e3}
\end{equation}
Here,  $\varepsilon$ is the quasiparticle energy measured from the Fermi
energy $E_F\equiv \hbar^2 k_F^2/2 m$,
${\cal H}_0\equiv{\bf p}^2/2 m + V - E_F$
is the single-particle Hamiltonian, with
$V({\bf r})$ the potential energy, $h({\bf r})$ is the exchange energy,
and $\Delta({\bf r})$ is the pair potential.
For simplicity, it is assumed that
the ferromagnet and the superconductor have identical ${\cal H}_0$.
For comparison with experiment, our model can easily be
extended to include differences in effective mass and band bottom.
We adopt the usual step-function model for the pair potential
\cite{btk82,lam91,tak92,bee92} and do
the same for the exchange energy \cite{bul82,kup86}.
Defining the FS interface at $x=0$ with
S at $x>0$, we have $\Delta({\bf r})=\Delta \Theta(x)$ and
$h({\bf r})=h_0 \Theta(- x)$, with $\Theta(x)$ the unit step-function.

A scattering formula for the linear-response
conductance of an NS junction is given
by Takane and Ebisawa
\cite{tak92}. Application to the FS case is straightforward,
\begin{equation}
G_{\text{FS}} = 2 \, \frac{e^2}{h}
\sum \limits_{\sigma=\uparrow,\downarrow} \mbox{Tr} \,
{\bf r}_{h\bar{\sigma},e\sigma}^\dagger
{\bf r}_{h\bar{\sigma},e\sigma}^{\vphantom{\dagger}}
 \: ,
\label{e6}
\end{equation}
where the matrix ${\bf r}_{h\bar{\sigma},e\sigma}$
contains the reflection amplitudes from incoming
electron modes with spin $\sigma$ to
outgoing hole modes with spin $\bar{\sigma}$
(opposite to $\sigma$) evaluated at the Fermi level
($\varepsilon=0$).
We first consider a ballistic point contact.
We assume that the dimensions of the contact are much greater than the
Fermi wavelength, as is appropriate for a metal, so that quantization
effects can be neglected. The number $N_\downarrow$ of minority spin modes
in the point contact (with area $\Omega$) is
$N_\downarrow =N_0 ( 1 - h_0/E_F)$, with $N_0 \equiv  k_F^2 \Omega / 4 \pi$
 the number of modes per spin for a non-ferromagnetic
($h_0=0$) contact of equal area.
The reflection matrices for this case can be evaluated
by matching the bulk solutions
for the ferromagnet and for the superconductor at the interface.
An incoming electron from the ferromagnet is either normally reflected as
an electron of the same spin or Andreev reflected
as a hole with the opposite spin.
(Transmission into the superconductor is not possible at
$\varepsilon=0$.)
The reflection matrices are diagonal, with elements
\begin{subequations}
\label{e4}
\begin{eqnarray}
r_{ee}&\equiv&
r_{e\sigma,e\sigma}=r_{h\sigma,h\sigma}=
\frac{k_\uparrow k_\downarrow - q^2}{k_\uparrow k_\downarrow + q^2}
\: ,
\label{e4a} \\
 r_{he} &\equiv&
r_{h\bar{\sigma},e\sigma}= r_{e\bar{\sigma},h\sigma} =
\frac{- 2 i q \sqrt{k_\uparrow k_\downarrow} }{k_\uparrow k_\downarrow + q^2}
\: ,
\label{e4b}
\end{eqnarray}
\end{subequations}%
where
the longitudinal wavevectors $k_{\uparrow(\downarrow)}$ in
the ferromagnet and $q$ in the superconductor are defined in
terms of the energy $E_n$ of the $n$-th transverse mode by
\begin{subequations}
\label{e5}
\begin{eqnarray}
q&=&\sqrt{ (2m/\hbar^2) ( E_F - E_n ) } \: ,
\label{e5a} \\
k_{\uparrow} &=&
\sqrt{ (2m/\hbar^2) ( E_F - E_n +  h_0) } \: ,
\label{e5b}\\
k_{\downarrow} &=&
\sqrt{ (2m/\hbar^2) ( E_F - E_n - h_0) } \: .
\label{e5c}
\end{eqnarray}
\end{subequations}%
In the above expressions terms of order $\Delta/E_F$ are neglected
\cite{notAA}.
Note that $|r_{ee}|^2 + |r_{he}|^2
= 1$, as required from quasiparticle conservation.
It follows from
Eq.\ (\ref{e4}) that a clean FS junction does not exhibit
complete Andreev reflection,
in contrast to the NS case.
This is due to the potential step  the particle
passes when being Andreev reflected to the opposite spin band.

\begin{figure}[tb]
\centerline{\includegraphics[width=0.9\linewidth]{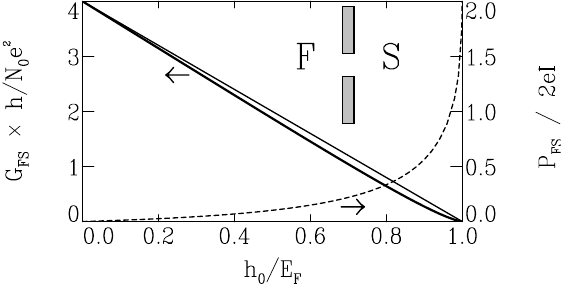}}
\caption{
The conductance $G_{\text{FS}}$ (full curves)
and the shot-noise power $P_{\text{FS}}$
(dashed)
of a ballistic point contact in a ferromagnet--superconductor junction
(see inset),
as a function of the exchange energy
$h_0$.
The thick line represents the exact result (\protect\ref{e7})
for $G_{\text{FS}}$,
the thin line the estimation (\protect\ref{e2}).
}
\label{f1}
\end{figure}

Because of the large number of modes the trace in Eq.\ (\ref{e6})
can be replaced by an integration, which can be evaluated analytically.
The result is
\begin{eqnarray}
G_{\text{FS}}
&=& 4 \,  \frac{e^2}{h} \, N_0 \,
\frac{4 }{15 \eta^4} \,
\times \nonumber \\
& &
[ \sqrt{1-\eta^2} (6 - 7 \eta^2 + \eta^4) - 6 + 10 \eta^2 - 4
\eta^5 ] \: ,
\label{e7}
\end{eqnarray}
where $\eta \equiv h_0/E_F$.
The conductance is plotted in Fig.\ \ref{f1},
and compared with the
semiclassical estimate
from Eq.\ (\ref{e2}), which turns out to be
quite accurate.
Since $N_\uparrow + N_\downarrow = 2 N_0$ one has from Eq.\ (\ref{e1})
$G_{\text{FN}} > G_{\text{FS}}$ if $h_0 > 0.47 E_F$, or
equivalently $N_\downarrow / N_\uparrow < 0.36$.

Further information on the Andreev reflection at the FS interface can
be obtained from
the shot-noise power $P$ of the junction.
Shot noise is the time-dependent fluctuation in the current
due to the discreteness of the charges.
For uncorrelated electron transmission,
one has the maximal noise power of
a Poisson process $P_{\text{Poisson}} \equiv 2 e I$,
with $I$ the mean current.
On the one hand,
correlations due to the Pauli principle reduce $P$
below $P_{\text{Poisson}}$ \cite{khl87,les89}.
On the other hand, Cooper-pair transport across an NS junction
has been shown to manifest itself  as a doubling of the
maximal noise power \cite{khl87,jon94}.
We apply the general result of Ref.\ \cite{jon94}
to the FS junction
\begin{equation}
P_{\text{FS}} = \frac{8 e^3 V}{h}
\sum \limits_{\sigma=\uparrow,\downarrow} \mbox{Tr} \,
{\bf r}_{h\bar{\sigma},e\sigma}^\dagger
{\bf r}_{h\bar{\sigma},e\sigma}^{\vphantom{\dagger}}
( {\bf 1} -
{\bf r}_{h\bar{\sigma},e\sigma}^\dagger
{\bf r}_{h\bar{\sigma},e\sigma}^{\vphantom{\dagger}}
)
\: .
\label{e8}
\end{equation}
Substitution of Eq.\ (\ref{e4b}) into Eq.\ (\ref{e8})
yields the shot-noise power of a ballistic point contact,
plotted in Fig.\ \ref{f1}.
The shot noise increases from complete suppression for a non-ferromagnetic
($h_0=0$) junction to twice the Poisson noise for a
half-metallic ferromagnet ($h_0=E_F$).
The initial increase is slow, indicating that the $N_\downarrow$ modes
undergo nearly complete Andreev reflection.
However, for higher exchange energies the Andreev reflection
probability decreases in favour of the normal reflection probability.
This is manifested by the increase in the shot-noise power.

\begin{figure}[tb]
\centerline{\includegraphics[width=0.9\linewidth]{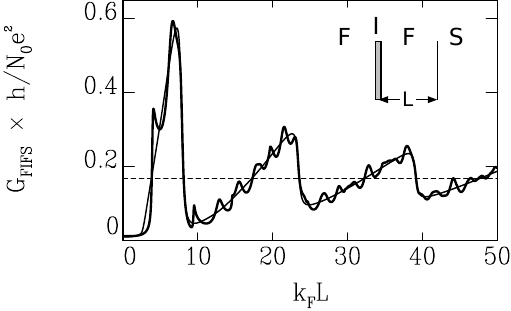}}
\caption{
The conductance $G_{\text{FIFS}}$ of a clean FIFS junction
containing a planar tunnel barrier (transparency $\Gamma$)
on the ferromagnetic side,
as a function of the separation $L$ from the
interface (see inset).
The thick solid line is computed from Eq.\ (\protect\ref{e11}) for
$\Gamma=0.1$, $h_0=0.2 E_F$.
For the thin line normal reflection at the FS interface
is neglected ($r_{ee}=0$). The dashed line is the classical
large-$L$ limit.}
\label{f2}
\end{figure}

\begin{figure}[tb]
\centerline{\includegraphics[width=0.9\linewidth]{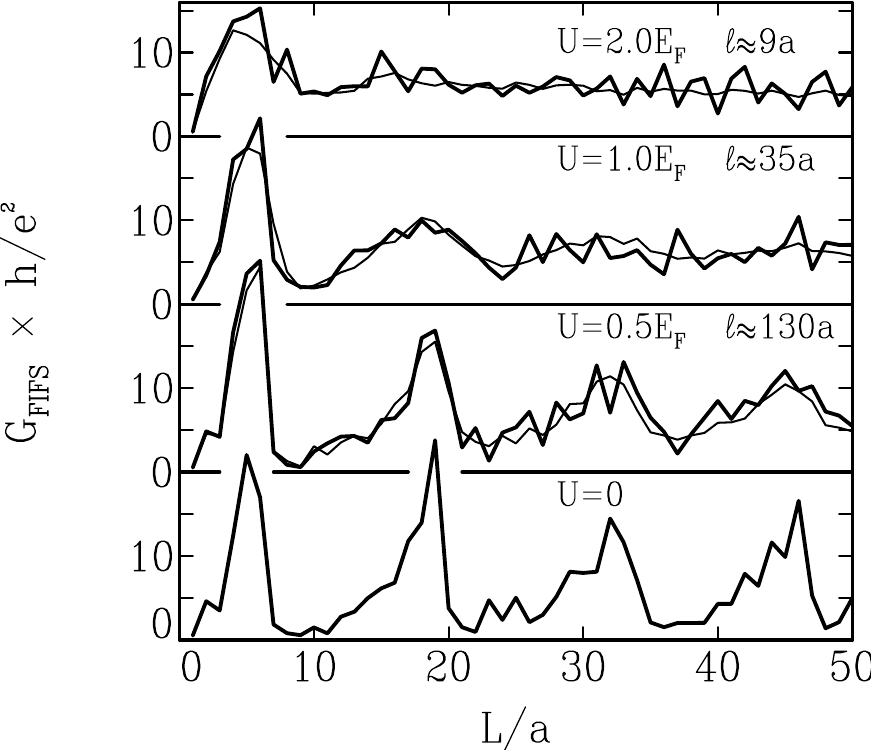}}
\caption{
Numerical calculation of the effect of disorder in the ferromagnet
on the oscillations shown in Fig.\ \protect\ref{f2} for a clean
junction.
The disordered region is modeled by a $L \times W$ square lattice
(lattice constant $a$) with random on-site disorder
(uniformly distributed between $\pm U/2$).
The width $W=101a$ is fixed
and the length $L$ is varied on the horizontal axis.
The results shown are for $E_F=\hbar^2/2ma^2$, $h_0=0.2 E_F$, $\Gamma=0.1$,
and for various $U$.
For each disorder strength $U$ the bulk mean free path $\ell$
is given.
Thick lines belong to one realization of disorder, thin to
an average over 20 realizations.
}
\label{f3}
\end{figure}

The second system we consider is an FIFS junction
which contains a planar tunnel barrier (I) at $x=-L$.
The barrier is modeled by a channel- and spin-independent
transmission probability $\Gamma \in [0,1]$.
The matrix
${\bf r}_{h\bar{\sigma},e\sigma}^\dagger
{\bf r}_{h\bar{\sigma},e\sigma}^{\vphantom{\dagger}}$
in Eq.\ (\ref{e6}) is diagonal, with elements
\begin{eqnarray}
| r_{h\bar{\sigma},e\sigma} |^2 =  \Gamma^2 \, |r_{he}|^2
&\Bigl\{ & 1 +  2 \rho^2 \cos(\chi_\uparrow-\chi_\downarrow) + \rho^4
\nonumber \\
&+&
2 r_{ee} \rho (1+\rho^2) ( \cos \chi_\uparrow + \cos \chi_\downarrow )
\nonumber \\
&+&
2 r_{ee}^2 \rho^2 [ 1 + \cos(\chi_\uparrow+\chi_\downarrow)]
\, \Bigl\}^{-1}
\, ,
\label{e11}
\end{eqnarray}
where $\rho\equiv\sqrt{1-\Gamma}$ and $\chi_\sigma\equiv2 k_\sigma L$.
Eq.\ (\ref{e11}) describes resonant Andreev reflection: Due to the
different wavevector of up electrons and down holes,
$| r_{h\bar{\sigma},e\sigma} |^2$ varies
as a function of $\chi_\uparrow$ and $\chi_\downarrow$ between
$\Gamma^2$, the value for a two-particle tunneling process, and
1 for full resonance. The conductance $G_{\text{FIFS}}$ is evaluated by
substitution of Eq.\ (\ref{e11}) into Eq.\ (\ref{e6}). It is
depicted in Fig.\ \ref{f2} as a function of $L$ for
$h_0=0.2 E_{\text{F}}$ and $\Gamma=0.1$.
The resonances have a dominant
period $\delta L=\pi\hbar v_F/2 h_0 (=5 \pi
k_F^{-1}$ in Fig.\ \ref{f2}), which
is caused by the simplest round-trip containing two Andreev reflections
and two barrier reflections. Superimposed one sees
oscillations with smaller period, caused by longer trajectories in
which also normal reflections at the FS interface occur.
This becomes clear when we calculate $G_{\text{FS}}$
with $r_{ee}$ set to zero, which is also shown in Fig.\
\ref{f2}. For large $L$, $G_{\text{FIFS}}$ approaches the
classical (i.e.\ all interferences are neglected)
value $4 (e^2/h) N_\downarrow \Gamma / (2 - \Gamma)$.
The oscillations in Fig.\ \ref{f2} are distinct from the
Tomasch oscillations known to occur in the
{\em non-linear} differential conductance of NINS junctions
\cite{tom65}.
There, quasi-bound states arise because electron and hole wavevectors disperse
if $\varepsilon > 0$. However, in linear response
$G_{\text{NINS}}=4 (e^2/h) N_0 \Gamma^2/(2-\Gamma)^2$,
independent of $L$ \cite{btk82}.
In the ferromagnetic junction the resonances do not vanish in linear
response, in contrast to the Tomasch oscillations. The quasi-bound states
at the Fermi level are a direct consequence of
the change in spin band upon Andreev reflection.

We believe that both phenomena are experimentally accessible.
The FS point contact can be constructed according to the
nanofabrication technique of Ref.\ \cite{hol91}.
The FIFS junction can be made
by growing a wedge-shaped layer of ferromagnet on a
superconducting
substrate and then depositing a thin oxide layer.
This allows a measurement of $G_{\text{FIFS}}$
for different values of $L$. It is not necessary for the
contact area to be small, so that
no nanofabrication techniques are needed.
(Note, that in order to observe the resonances due to the
quasi-bound states it is not essential that the contact on top
of the barrier is a ferromagnet.)
To estimate the effect of disorder (growth imperfections and impurities)
on the resonances, we have numerically calculated $G_{\text{FIFS}}$ for
a disordered ferromagnet between the barrier and the FS interface.
The computations are similar
to the NS case treated in Ref.\ \cite{mar93}.
The disordered region is modeled by a tight-binding Hamiltonian
on a square lattice with a random impurity
potential at each site.
(For computational efficiency the geometry is two-dimensional,
but this makes no qualitative difference.)
The matrix ${\bf r}_{h\bar{\sigma},e\sigma}$ is
obtained by combining the scattering matrix of the disordered region
with the reflection coefficients for the FS interface (\ref{e4}).
We then calculate $G_{\text{FS}}$ through Eq.\ (\ref{e6}).
The result for various disorder strengths
is shown in Fig.\ \ref{f3}. For the clean case we
recognize a behavior similar to Fig.\ \ref{f2}. Adding some disorder
removes the small-period oscillations but preserves the dominant
oscillations.
Only quite a strong disorder (for the top curve
$k_F$ $\times$ bulk mean free path $\simeq$ 9)
is able to smooth away the resonances.

In summary, we have shown that the transport properties of
ferromagnet--superconductor junctions are qualitatively different from
the non-ferromagnetic case,
because the
Andreev reflection is modified by the exchange interaction in the ferromagnet.
Two illustrative examples
have been given:
For a ballistic FS point contact it is found that the conductance can be
both larger or smaller than the normal-state value and
for an FIFS junction containing a tunnel barrier conductance
resonances are predicted to occur in linear response.

We are especially grateful to H. van Houten for suggesting the problem
treated in this paper. Furthermore, we thank P. J. Kelly and C. M. Schep
for useful discussions.
This research was supported
by the Dutch Science Foundation NWO/FOM.


\begin{references}
\bibitem{notAA}
This is the Andreev approximation \cite{and64}.
One can easily go beyond it by including terms of order $\Delta/E_F$
in Eqs.\ (\ref{e4}) and (\ref{e5}).
We have checked that this
has only a small influence on our final results.
In fact, the larger $h_0$, the more accurate
is the Andreev approximation.
\end{references}
\end{document}